\documentclass[aps,prl,showpacs,twocolumn,superscriptaddress]{revtex4}  

\usepackage{graphicx}
\usepackage{dcolumn}
\usepackage{bm}
\usepackage{amssymb}   
\usepackage{epsfig}

\usepackage{bbm}
\usepackage{url}
\newcommand{\fb} {{\rm fb}^{-1}}

\newcommand{\lumi} {6.8~\fb}

\newcommand{\GeV}{{\rm GeV}}
\newcommand{\GeVc}{{\rm GeV/}c}
\newcommand{\GeVcsq}{{\rm GeV/}c^2}

\newcommand{\MeVcsq}{{\rm MeV/}c^2}

\newcommand{\kmmRelBr}{0.46}
\newcommand{\kmmRelBrStat}{0.04}
\newcommand{\kmmRelBrSyst}{0.02}
\newcommand{\kstmmRelBr}{0.77}
\newcommand{\kstmmRelBrStat}{0.08}
\newcommand{\kstmmRelBrSyst}{0.03}
\newcommand{\phimmRelBr}{1.13}
\newcommand{\phimmRelBrStat}{0.19}
\newcommand{\phimmRelBrSyst}{0.07}
\newcommand{\ksmmRelBr}{0.37}
\newcommand{\ksmmRelBrStat}{0.12}
\newcommand{\ksmmRelBrSyst}{0.02}
\newcommand{\kstpmmRelBr}{0.67}
\newcommand{\kstpmmRelBrStat}{0.22}
\newcommand{\kstpmmRelBrSyst}{0.04}
\newcommand{\lmmmRelBr}{2.45}
\newcommand{\lmmmRelBrStat}{0.59}
\newcommand{\lmmmRelBrSyst}{0.29}

\newcommand{\kmmBr}{0.46}
\newcommand{\kmmBrStat}{0.04}
\newcommand{\kmmBrSyst}{0.02}
\newcommand{\kstmmBr}{1.02}
\newcommand{\kstmmBrStat}{0.10}
\newcommand{\kstmmBrSyst}{0.06}
\newcommand{\phimmBr}{1.47}
\newcommand{\phimmBrStat}{0.24}
\newcommand{\phimmBrSyst}{0.46}
\newcommand{\ksmmBr}{0.32}
\newcommand{\ksmmBrStat}{0.10}
\newcommand{\ksmmBrSyst}{0.02}
\newcommand{\kstpmmBr}{0.95}
\newcommand{\kstpmmBrStat}{0.32}
\newcommand{\kstpmmBrSyst}{0.08}
\newcommand{\lmmmBr}{1.73}
\newcommand{\lmmmBrStat}{0.42}
\newcommand{\lmmmBrSyst}{0.55}

\newcommand{\kallmmBr}{0.42}
\newcommand{\kallmmBrStat}{0.04}
\newcommand{\kallmmBrSyst}{0.02}
\newcommand{\kstallmmBr}{1.01}
\newcommand{\kstallmmBrStat}{0.10}
\newcommand{\kstallmmBrSyst}{0.05}

\newcommand{\bz}{B^0}

\newcommand{\bp}{B^+}

\newcommand{\bs}{B^0_s}

\newcommand{\lb}{\Lambda^0_b}

\newcommand {\kst}{K^{*0}}
\newcommand {\kstp}{K^{*+}}
\newcommand {\kstall}{K^{*}}
\newcommand {\ks}{K_S^{0}}
\newcommand {\kz}{K^{0}}
\newcommand {\lm}{\Lambda}

\newcommand {\kstpdg}{K^{*}(892)^{0}}
\newcommand {\kstppdg}{K^{*}(892)^{+}}

\newcommand {\phipdg}{\phi(1020)}

\newcommand{\kmm}{K^+ \mu^+ \mu^-}
\newcommand{\kallmm}{K \mu^+ \mu^-}
\newcommand{\ksmm}{\ks \mu^+ \mu^-}
\newcommand{\kzmm}{\kz \mu^+ \mu^-}
\newcommand{\kstmm}{\kst \mu^+ \mu^-}
\newcommand{\kstpmm}{\kstp \mu^+ \mu^-}
\newcommand{\kstallmm}{\kstall \mu^+ \mu^-}
\newcommand{\phimm}{\phi \mu^+ \mu^-}
\newcommand{\lmmm}{\lm \mu^+ \mu^-}
\newcommand{\hmm}{h \mu^+ \mu^-}

\newcommand{\smm}{s \mu^+ \mu^-}

\newcommand{\bsmm}{b \to \smm}

\newcommand{\bkallmm}{B \to \kallmm}
\newcommand{\bkstallmm}{B \to \kstallmm}
\newcommand{\bkmmincl}{B \to K^{(*)}\mu^+\mu^-}

\newcommand{\jpsi}{J/{\small \psi}}

\newcommand{\psiphi}{\jpsi \phi}

\newcommand{\psilm}{\jpsi \lm}
\newcommand{\psih}{\jpsi h}

\newcommand{\bpkmm}{\bp \to \kmm}
\newcommand{\bzksmm}{\bz \to \ksmm}
\newcommand{\bzkzmm}{\bz \to \kzmm}
\newcommand{\bzkstmm}{\bz \to \kstmm}
\newcommand{\bpkstpmm}{\bp \to \kstpmm}
\newcommand{\bsphimm}{\bs \to \phimm}
\newcommand{\lblmmm}{\lb \to \lmmm}
\newcommand{\bhmm}{H_b \to \hmm}

\newcommand{\bzkstmmpdg}{\bz \to \kstpdg \mu^+ \mu^-}
\newcommand{\bpkstpmmpdg}{\bp \to \kstppdg \mu^+ \mu^-}
\newcommand{\bsphimmpdg}{\bs \to \phipdg \mu^+ \mu^-}

\newcommand{\bspsiphi}{\bs \to \psiphi}

\newcommand{\lbpsilm}{\lb \to \psilm}
\newcommand{\bpsih}{H_b \to \psih}

\newcommand{\bpsikincl}{B \to \jpsi K^{(*)}}

\newcommand{\psimm}{\jpsi \to \mu^+ \mu^-}

\newcommand{\Mmm}{M_{\mu\mu}}

\newcommand{\br}{{\mathcal{B}}}

\newcommand{\BaBar}{{\mbox{\sl B\hspace{-0.4em} {\small\sl A} \hspace{-0.37em}\sl B \hspace{-0.4em}}}{\small\sl A\hspace{-0.02em}R}}

\begin{document}

\title{\bf \boldmath 
Observation of the Baryonic Flavor-Changing Neutral Current Decay $\lblmmm$ }

\affiliation{Institute of Physics, Academia Sinica, Taipei, Taiwan 11529, Republic of China} 
\affiliation{Argonne National Laboratory, Argonne, Illinois 60439, USA} 
\affiliation{University of Athens, 157 71 Athens, Greece} 
\affiliation{Institut de Fisica d'Altes Energies, Universitat Autonoma de Barcelona, E-08193, Bellaterra (Barcelona), Spain} 
\affiliation{Baylor University, Waco, Texas 76798, USA} 
\affiliation{Istituto Nazionale di Fisica Nucleare Bologna, $^{bb}$University of Bologna, I-40127 Bologna, Italy} 
\affiliation{Brandeis University, Waltham, Massachusetts 02254, USA} 
\affiliation{University of California, Davis, Davis, California 95616, USA} 
\affiliation{University of California, Los Angeles, Los Angeles, California 90024, USA} 
\affiliation{Instituto de Fisica de Cantabria, CSIC-University of Cantabria, 39005 Santander, Spain} 
\affiliation{Carnegie Mellon University, Pittsburgh, Pennsylvania 15213, USA} 
\affiliation{Enrico Fermi Institute, University of Chicago, Chicago, Illinois 60637, USA}
\affiliation{Comenius University, 842 48 Bratislava, Slovakia; Institute of Experimental Physics, 040 01 Kosice, Slovakia} 
\affiliation{Joint Institute for Nuclear Research, RU-141980 Dubna, Russia} 
\affiliation{Duke University, Durham, North Carolina 27708, USA} 
\affiliation{Fermi National Accelerator Laboratory, Batavia, Illinois 60510, USA} 
\affiliation{University of Florida, Gainesville, Florida 32611, USA} 
\affiliation{Laboratori Nazionali di Frascati, Istituto Nazionale di Fisica Nucleare, I-00044 Frascati, Italy} 
\affiliation{University of Geneva, CH-1211 Geneva 4, Switzerland} 
\affiliation{Glasgow University, Glasgow G12 8QQ, United Kingdom} 
\affiliation{Harvard University, Cambridge, Massachusetts 02138, USA} 
\affiliation{Division of High Energy Physics, Department of Physics, University of Helsinki and Helsinki Institute of Physics, FIN-00014, Helsinki, Finland} 
\affiliation{University of Illinois, Urbana, Illinois 61801, USA} 
\affiliation{The Johns Hopkins University, Baltimore, Maryland 21218, USA} 
\affiliation{Institut f\"{u}r Experimentelle Kernphysik, Karlsruhe Institute of Technology, D-76131 Karlsruhe, Germany} 
\affiliation{Center for High Energy Physics: Kyungpook National University, Daegu 702-701, Korea; Seoul National University, Seoul 151-742, Korea; Sungkyunkwan University, Suwon 440-746, Korea; Korea Institute of Science and Technology Information, Daejeon 305-806, Korea; Chonnam National University, Gwangju 500-757, Korea; Chonbuk National University, Jeonju 561-756, Korea} 
\affiliation{Ernest Orlando Lawrence Berkeley National Laboratory, Berkeley, California 94720, USA} 
\affiliation{University of Liverpool, Liverpool L69 7ZE, United Kingdom} 
\affiliation{University College London, London WC1E 6BT, United Kingdom} 
\affiliation{Centro de Investigaciones Energeticas Medioambientales y Tecnologicas, E-28040 Madrid, Spain} 
\affiliation{Massachusetts Institute of Technology, Cambridge, Massachusetts 02139, USA} 
\affiliation{Institute of Particle Physics: McGill University, Montr\'{e}al, Qu\'{e}bec, Canada H3A~2T8; Simon Fraser University, Burnaby, British Columbia, Canada V5A~1S6; University of Toronto, Toronto, Ontario, Canada M5S~1A7; and TRIUMF, Vancouver, British Columbia, Canada V6T~2A3} 
\affiliation{University of Michigan, Ann Arbor, Michigan 48109, USA} 
\affiliation{Michigan State University, East Lansing, Michigan 48824, USA}
\affiliation{Institution for Theoretical and Experimental Physics, ITEP, Moscow 117259, Russia}
\affiliation{University of New Mexico, Albuquerque, New Mexico 87131, USA} 
\affiliation{Northwestern University, Evanston, Illinois 60208, USA} 
\affiliation{The Ohio State University, Columbus, Ohio 43210, USA} 
\affiliation{Okayama University, Okayama 700-8530, Japan} 
\affiliation{Osaka City University, Osaka 588, Japan} 
\affiliation{University of Oxford, Oxford OX1 3RH, United Kingdom} 
\affiliation{Istituto Nazionale di Fisica Nucleare, Sezione di Padova-Trento, $^{cc}$University of Padova, I-35131 Padova, Italy} 
\affiliation{LPNHE, Universite Pierre et Marie Curie/IN2P3-CNRS, UMR7585, Paris, F-75252 France} 
\affiliation{University of Pennsylvania, Philadelphia, Pennsylvania 19104, USA}
\affiliation{Istituto Nazionale di Fisica Nucleare Pisa, $^{dd}$University of Pisa, $^{ee}$University of Siena and $^{ff}$Scuola Normale Superiore, I-56127 Pisa, Italy} 
\affiliation{University of Pittsburgh, Pittsburgh, Pennsylvania 15260, USA} 
\affiliation{Purdue University, West Lafayette, Indiana 47907, USA} 
\affiliation{University of Rochester, Rochester, New York 14627, USA} 
\affiliation{The Rockefeller University, New York, New York 10065, USA} 
\affiliation{Istituto Nazionale di Fisica Nucleare, Sezione di Roma 1, $^{gg}$Sapienza Universit\`{a} di Roma, I-00185 Roma, Italy} 

\affiliation{Rutgers University, Piscataway, New Jersey 08855, USA} 
\affiliation{Texas A\&M University, College Station, Texas 77843, USA} 
\affiliation{Istituto Nazionale di Fisica Nucleare Trieste/Udine, I-34100 Trieste, $^{hh}$University of Trieste/Udine, I-33100 Udine, Italy} 
\affiliation{University of Tsukuba, Tsukuba, Ibaraki 305, Japan} 
\affiliation{Tufts University, Medford, Massachusetts 02155, USA} 
\affiliation{Waseda University, Tokyo 169, Japan} 
\affiliation{Wayne State University, Detroit, Michigan 48201, USA} 
\affiliation{University of Wisconsin, Madison, Wisconsin 53706, USA} 
\affiliation{Yale University, New Haven, Connecticut 06520, USA} 
\author{T.~Aaltonen}
\affiliation{Division of High Energy Physics, Department of Physics, University of Helsinki and Helsinki Institute of Physics, FIN-00014, Helsinki, Finland}
\author{B.~\'{A}lvarez~Gonz\'{a}lez$^v$}
\affiliation{Instituto de Fisica de Cantabria, CSIC-University of Cantabria, 39005 Santander, Spain}
\author{S.~Amerio}
\affiliation{Istituto Nazionale di Fisica Nucleare, Sezione di Padova-Trento, $^{cc}$University of Padova, I-35131 Padova, Italy} 

\author{D.~Amidei}
\affiliation{University of Michigan, Ann Arbor, Michigan 48109, USA}
\author{A.~Anastassov}
\affiliation{Northwestern University, Evanston, Illinois 60208, USA}
\author{A.~Annovi}
\affiliation{Laboratori Nazionali di Frascati, Istituto Nazionale di Fisica Nucleare, I-00044 Frascati, Italy}
\author{J.~Antos}
\affiliation{Comenius University, 842 48 Bratislava, Slovakia; Institute of Experimental Physics, 040 01 Kosice, Slovakia}
\author{G.~Apollinari}
\affiliation{Fermi National Accelerator Laboratory, Batavia, Illinois 60510, USA}
\author{J.A.~Appel}
\affiliation{Fermi National Accelerator Laboratory, Batavia, Illinois 60510, USA}
\author{A.~Apresyan}
\affiliation{Purdue University, West Lafayette, Indiana 47907, USA}
% ...
\collaboration{CDF Collaboration\footnote{With visitors from $^a$University of Massachusetts Amherst, Amherst, Massachusetts 01003,
$^b$Istituto Nazionale di Fisica Nucleare, Sezione di Cagliari, 09042 Monserrato (Cagliari), Italy,
$^c$University of California Irvine, Irvine, CA  92697, 
$^d$University of California Santa Barbara, Santa Barbara, CA 93106
$^e$University of California Santa Cruz, Santa Cruz, CA  95064,
$^f$CERN,CH-1211 Geneva, Switzerland,
$^g$Cornell University, Ithaca, NY  14853, 
$^h$University of Cyprus, Nicosia CY-1678, Cyprus, 
$^i$University College Dublin, Dublin 4, Ireland,
$^j$University of Fukui, Fukui City, Fukui Prefecture, Japan 910-0017,
$^k$Universidad Iberoamericana, Mexico D.F., Mexico,
$^l$Iowa State University, Ames, IA  50011,
$^m$University of Iowa, Iowa City, IA  52242,
$^n$Kinki University, Higashi-Osaka City, Japan 577-8502,
$^o$Kansas State University, Manhattan, KS 66506,
$^p$University of Manchester, Manchester M13 9PL, England,
$^q$Queen Mary, University of London, London, E1 4NS, England,
$^r$Muons, Inc., Batavia, IL 60510,
$^s$Nagasaki Institute of Applied Science, Nagasaki, Japan, 
$^t$National Research Nuclear University, Moscow, Russia,
$^u$University of Notre Dame, Notre Dame, IN 46556,
$^v$Universidad de Oviedo, E-33007 Oviedo, Spain, 
$^w$Texas Tech University, Lubbock, TX  79609, 
$^x$IFIC(CSIC-Universitat de Valencia), 56071 Valencia, Spain,
$^y$Universidad Tecnica Federico Santa Maria, 110v Valparaiso, Chile,
$^z$University of Virginia, Charlottesville, VA  22906,
$^{aa}$Yarmouk University, Irbid 211-63, Jordan,
$^{ii}$On leave from J.~Stefan Institute, Ljubljana, Slovenia, 
}}
\noaffiliation


\begin{abstract}

We report the  first  observation of the baryonic flavor-changing neutral current decay $\lblmmm$ with $24$  signal events 
and a statistical significance  of 5.8 Gaussian standard deviations.
This measurement uses a $p\bar{p}$ collisions data sample corresponding to $\lumi$    at $\sqrt{s}=1.96~{\rm TeV}$  collected 
by the CDF~II detector  at the Tevatron collider.
The total and differential branching ratios for $\lblmmm$ 
are measured. We find $\br(\lblmmm) = [\lmmmBr \pm \lmmmBrStat({\rm stat}) \pm \lmmmBrSyst({\rm syst})] \times 10^{-6}$. 
We also report the first measurement of the differential branching ratio of $\bsphimm$ using $49$ signal events.  
In addition, we report branching ratios for $\bpkmm$, $\bzkzmm$ and $B \to K^*(892) \mu^+ \mu^-$ decays.

\end{abstract}

\maketitle


Rare decays of hadrons containing bottom quarks through the process $\bsmm$, 
where $b$ is a bottom quark and $s$ is a strange quark, occur in the standard model (SM) 
with $\mathcal{O}(10^{-6})$ branching ratios~\cite{Melikhov:1997wp_APS, Ali:1999mm_APS}.
The $b$ and $s$ quarks carry the same charge but different flavor, 
so this process is a flavor-changing neutral-current (FCNC) decay.
FCNC decays are suppressed at tree level in the SM, and must occur through higher order,
 and more suppressed, loop amplitudes.
Their suppressed nature and clean experimental signature, 
along with reliable theoretical  predictions for their rates~\cite{Ali:1999mm_APS,  Wang:2008ni_APS, Chang:2011jka_APS},
make them excellent search channels for new physics.
With multibody final states, these decays offer sensitivity to new physics 
in a number of kinematic distributions in addition to the total branching ratio.
In this Letter we report measurements of the total branching ratios of FCNC decays,
as well as their differential branching ratios
as a function of $q^2 \equiv \Mmm^2 c^2$, where $\Mmm$ is the dimuon invariant mass.
Exclusive decays of $\bkmmincl$  have been observed by \BaBar~\cite{Aubert:2008ps}, Belle~\cite{Wei:2009zv},
and CDF~\cite{Aaltonen:2008xf_APS_Aaltonen:2011cn}. 
The CDF experiment also recently reported the  observation of $\bsphimmpdg$~\cite{Aaltonen:2008xf_APS_Aaltonen:2011cn}.
No significant departure from the SM has been found thus far.

In addition, the study of the baryonic $\bsmm$ decays is very important,
since the baryonic FCNC decays are sensitive to the helicity structure of effective Hamiltonian which is lost in the hadronization of the mesonic decays~\cite{Wang:2008sm_APS}.
Although the theoretical calculations of the exclusive baryonic $\bsmm$ decays have large uncertainties compared to the mesonic decays due to additional degrees of freedom in the baryon bound states, the measurements of the total and the differential branching ratios can help the improvement of the theoretical treatments.
One can also compare the measurements of the mesonic $\bsmm$ decays with the baryonic decays,
which follow the common quark transition.
Measurements of both mesonic and baryonic FCNC decays therefore 
provide additional tests of the SM and its extensions.
However, no $b$ baryon FCNC decay has been observed and
there are few experimental constraints on their decay rates.
The $\lblmmm$ decay is considered 
promising in this respect~\cite{Chen:2001zc_APS,Wang:2008sm_APS,Aslam:2008hp_APS, Aliev:2010uy_APS} and
experimentally accessible since
the branching ratio is predicted as $(4.0\pm1.2)\times 10^{-6}$~\cite{Aliev:2010uy_APS}.

The data sample used in the measurements reported in this Letter  corresponds  to an integrated luminosity
of $\lumi$ from $p\bar{p}$ collisions at a center-of-mass energy of $\sqrt{s}=1.96~{\rm TeV}$
collected with the CDF II detector between March 2002 and June 2010.
The $\lblmmm$ decay is reconstructed and measurements are made of the total 
branching ratio and the differential branching ratio as a function of $q^2$.
Besides the updated
branching ratios of $\bsphimm$, $\bpkmm$, and $\bzkstmmpdg$,
we report
the branching ratios of $\bzkzmm$ and $\bpkstpmmpdg$ which are measured for the first time in hadron collisions.
We also report the first measurement  of the 
differential branching ratio as a function of $q^2$ of $\bsphimm$.
To cancel the dominant systematic uncertainties, decay rates for each rare channel $\bhmm$ are measured relative to the corresponding resonant channel $\bpsih$ with $\jpsi \to \mu^+ \mu^-$, used as a normalization,
where $H_b$ represents the $b$ hadron and
 $h$ stands for $\lm$, $\phi$, $K^+$, $\ks$, $\kst$, and $\kstp$.
Charge-conjugation is implied throughout the Letter.

The reconstruction of the exclusive $b$ hadron events starts with a
dimuon sample selected by the online trigger system~\cite{CDF:Trigger}  of  the CDF II
detector~\cite{Acosta:2004yw_APS}.  
The trigger system utilizes information from muon detectors 
and the central outer tracker~\cite{Aaltonen:2009tz_ref}.
Muon chambers CMU and CMX~\cite{Ascoli:1987av_Dorigo:2000ip_APS}
cover $|\eta| < 0.6$ and $0.6 < |\eta| < 1.0$,  respectively,~\cite{CDF_coordinate}.
The CMP muon chamber covers $|\eta|<0.6$ and is located behind the CMU and an additional steel absorber. 
The dimuon trigger requires a pair of oppositely charged particles
with a momentum transverse to the beam line
 $p_T\geq1.5\,\GeVc$, which
are matched to track segments in the CMU or CMX chambers.
At least one of the muon tracks is required to
have a CMU track segment.
The trigger also requires that the dimuon pair satisfies either
$L_{xy}>100~\mu \rm m$, where the transverse  decay length  $L_{xy}$ is 
the flight distance between the dimuon vertex and the event primary vertex~\cite{CDF:SVT},
or $p_T > 3.0\,\GeVc$ and matched segments in both CMU and CMP chambers for one of the muon candidates.

Offline event selection starts with the triggered dimuon pairs.
Each offline track is required to satisfy more stringent requirements on the number of hits used to reconstruct the track.
The dimuon selection requirements used in the trigger are repeated with the higher quality offline tracks.
The decay length and 
invariant mass of each dimuon pair are calculated after a vertex fit
using the muon tracks. 
Dimuon pairs are classified according to their invariant mass $\Mmm$.
Dimuons  from FCNC $b$ hadron decays are 
required  to be inconsistent with decaying 
from $\jpsi$ ($\psi(2S)$) mesons by requiring $q^2$ values
 outside the window of
$8.68$ $(12.86)$ $<q^2<$ $10.09$ $(14.18)$
$\,\GeV^2/c^2$~\cite{Aaltonen:2008xf_APS_Aaltonen:2011cn}.
The $\jpsi$ candidates  are 
required to have $\Mmm$ within $50\,\MeVcsq$ of the known $\jpsi$ mass~\cite{Nakamura:2010zzi_APS}.

The $\lblmmm$ candidates are  selected by combining the dimuon
pairs with $\lm$ baryons reconstructed from decays $\lm \to p \pi^-$.
The  $p \pi^-$ pairs are required to have invariant mass consistent
with the known $\lm$ mass~\cite{Nakamura:2010zzi_APS}, $p_T \geq 1.0\,\GeVc$,
and a vertex displaced from the dimuon vertex.  
The transverse momentum of the $\lb$ candidate is required
to be greater than $4.0\,\GeVc$.    
Candidates with an invariant mass calculated from two or three daughter
particles compatible with $\jpsi$, $\psi(2S)$, $D^0$, $D^+$, $D_s^+$, or $\Lambda_c$ masses are 
rejected to remove backgrounds from these charm-hadron decays~\cite{Aaltonen:2008xf_APS_Aaltonen:2011cn}.
The $\bsphimm$ candidates  are reconstructed from dimuons together with
a pair of oppositely-charge kaons consistent with a $\phi$ decay
with a selection similar to that of $\lblmmm$.
The $B^{0,+}\to {\cal K}^{0,+} \mu^+\mu^-$ candidates,
where ${\cal K}^{0,+}$ is one of $\{K^+, \ks, \kst, \kstp\}$,
are formed
from a dimuon combined with up to three 
charged tracks. 
The $\ks$ meson is reconstructed in its $\pi^+\pi^-$ final state by requiring the dipion mass
to be consistent with the known $\ks$ mass~\cite{Nakamura:2010zzi_APS}.
Details about the reconstruction of the decays of $\kst \to K^+\pi^-$ and $\phi \to K^+K^-$
can be found in Ref.~\cite{Aaltonen:2008xf_APS_Aaltonen:2011cn}.
Cross-feed between $\lblmmm$ and $\bzksmm$  is suppressed by
evaluating the momentum imbalance of $\Lambda$ and $\ks$ daughters~\cite{Armentros:1954}.
We utilize the correlation between invariant mass and the asymmetry $\alpha \equiv (q_L^+ - q_L^-)/(q_L^+ + q_L^-)$, 
where $q_L^{+(-)}$ is the longitudinal momentum of the positive (negative) decay product relative to the direction of the $\lm$ or $\ks$.
We reject candidates satisfy
$-0.26<-1.9M(\ks)+|\alpha|<-0.15$ for $\ksmm$ and
$4.73<3.6M(\lm)+|\alpha|<4.78$ for $\lmmm$.
We remove 76 (90)\% of the cross-feed while the signal loss is 11 (7)\%
for $\lmmm$ ($\ksmm$). 
A residual cross-feed contamination of 0.1\% (0.6\%) to the $\lmmm$ ($\ksmm$) signal is considered as a systematic uncertainty.
To further optimize the event selection, an artificial neural network
(NN) classifier is trained using simulated signal events
and background events
taken from 
$H_b$ mass sidebands ($0.1$--$0.36\,\GeVcsq$ far from the known $H_b$ mass)
in data.
Some kinematical distributions of the simulated signal, e.g., the transverse momentum of $b$ hadron,
and the energy depositions of muon candidates in
the electromagnetic and hadron calorimeters, 
are corrected using scale factors extracted by comparing simulation to data in the normalization channels.
We use 70\% of the sideband events for the training, and use the remaining events to check that the NN does not bias or over suppress the mass distribution.
The optimized NN threshold is determined to maximize the average expected significance of the branching ratio,
using many kinematic
observables including  transverse momentum, invariant mass, vertex fit
qualities and muon identification qualities~\cite{Aaltonen:2008xf_APS_Aaltonen:2011cn}.

\begin{figure}[t]
  \begin{center}
    \begin{tabular}{c}
      \resizebox{.49\textwidth}{!}{\includegraphics[clip]{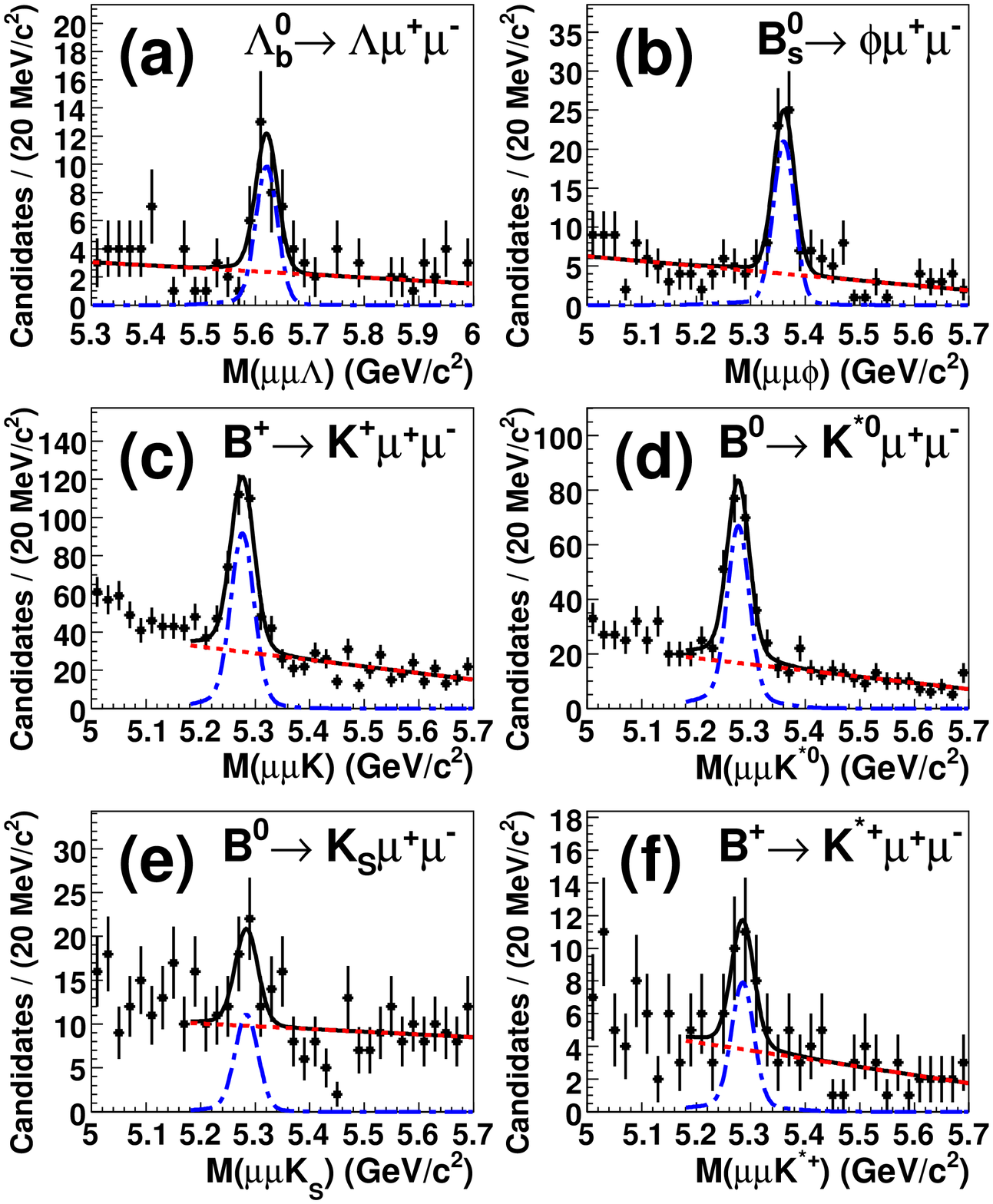}}
    \end{tabular}
  \end{center}
  \caption{
Invariant mass of (a) $\lblmmm$, (b) $\bsphimm$, (c) $\bpkmm$, (d) $\bzkstmm$, (e) $\bzksmm$, (f) $\bpkstpmm$, with fit results overlaid.
The histograms are the data. Solid, dashed-dotted, and dotted curves show the total fit, the signal PDF and the background PDF, respectively.
}
  \label{fig:rarebmass_data}
\end{figure}

The signal yield of the $\lblmmm$ candidates is obtained by an unbinned maximum likelihood fit
to the $\lb$ invariant mass distribution with the signal probability density
function (PDF) parametrized by Gaussian distributions using
simulated signals
and
the background PDF modeled by a linear function.
We fix the $\lb$ mass width for the rare decay while it is floated for the normalization channel.
Different mass width between data and the simulated signal is corrected by measured mass width ratio of the normalization channel between data and the simulated signal.
The signal region is defined within $\pm40$\,MeV/$c^2$ from the world average $\lb$ mass~\cite{Aaltonen:2008xf_APS_Aaltonen:2011cn}.
The statistical significance is obtained through a likelihood-ratio
test between the signal plus background and background-only hypotheses interpreted assuming it distributed as a $\chi^2$ variable.
The invariant mass distribution of the $\lblmmm$ candidates is shown in  Fig.~\ref{fig:rarebmass_data}.  
In the signal region, we observe $24\pm 5$ events from $\lblmmm$ decays while the total number of the signal candidates is 34.
The statistical significance of the signal $s$ corresponds to 5.8 Gaussian standard deviations.
The signal yields of $\bsphimm$ and other FCNC $B$ meson
decays are obtained by a similar procedure as that of $\lblmmm$.
Each channel uses independent NN weight and PDF.
The fit range for $\bp$ and $\bz$ decays is from $5.18$ to $5.70\,\GeVcsq$ to avoid the region of 
$5.0$--$5.18\,\GeVcsq$, which is dominated by 
the feed-down background from multibody decays of $b$ hadrons.
While the contribution from charmless $H_b$ decays is negligible due to the muon identification,
we estimate a $1\%$ crosstalk between $\bzkstmm$ and $\bsphimm$ using simulation, and correct for it.
Invariant mass distributions of $\bsphimm$ and other FCNC $B$ meson
decays are shown in Fig.~\ref{fig:rarebmass_data} and  signal
yields are listed in  Table~\ref{tab:summary_yield}.

\begin{table}[b]
  \caption{
    Summary of observed yields, the statistical significance $s$, and the relative efficiency $\varepsilon_{\rm rel}$.}
  \begin{center}
    \begin{tabular}{lcccc}
      \hline\hline
      \multicolumn{1}{c}{Mode} & $N_{\hmm}$ & $s$ $(\sigma)$ & $N_{\psih}$ & ~~~~$\varepsilon_{\rm rel}$\\
      \hline
     $\lblmmm$   & $24\pm5$    &  5.8 & $1740\pm50$ & $0.33\pm0.01$\\
      $\bsphimm$  & $49\pm7$   &  9.0 & $4560\pm80$ & $0.56\pm0.01$\\
      $\bpkmm$   & $234\pm19$  & 13.7 & $72200\pm300$ & $0.41\pm0.01$\\
      $\bzkstmm$  & $164\pm15$ & 13.7 & $28300\pm200$ & $0.45\pm0.02$\\
     $\bzksmm$   & $28\pm9$    &  3.5 & $9470\pm90$ & $0.47\pm0.01$\\
      $\bpkstpmm$ & $20\pm6$   &  3.5 & $4560\pm80$ & $0.38\pm0.02$\\
      \hline\hline 
    \end{tabular}
  \end{center}
  \label{tab:summary_yield}
\end{table}

The branching ratios of $\lblmmm$, $\bsphimm$, and $\bkmmincl$
are calculated by comparing their signal event yield to that of the normalization decay modes 
$\lbpsilm$, $\bspsiphi$, and $\bpsikincl$,
where $\jpsi \to \mu^+\mu^-$,
after the reconstruction efficiency correction:
\begin{equation}
\label{eq:br}
 \frac{\br(\bhmm)}{\br(\bpsih)}=
\frac{N_{\hmm}}{N_{\psih}}
\times \frac{\br(\psimm)}{\varepsilon_{\rm rel}},
\end{equation}
where
$N_{\hmm}$ is the $\hmm$ yield, $N_{\psih}$ is the $\psih$ yield  
for the normalization channel, and $\varepsilon_{\rm rel}\equiv{\varepsilon_{\hmm}}/{\varepsilon_{\psih}}$ is the relative
reconstruction efficiency determined from the simulation.
The calculated relative and absolute branching ratios are listed
in Table~\ref{tab:summary_br}.
The absolute branching ratios are obtained using world averages
of the $\psih$ decay rates~\cite{Nakamura:2010zzi_APS}.
The branching ratios of $\bzkzmm$ and $\bpkstpmm$ are measured for the first time in hadron collisions.

The dominant sources of systematic uncertainty are
the scale-factor reweighting of the simulated signal (the trigger efficiency near the threshold)
which ranges from $0.5\%$ to $4.0\%$ ($0.8\%$ to $7.2\%$), depending on the channel.
We estimate the former uncertainty from the comparison of the relative efficiencies with and without reweighting and
the latter uncertainty from the different $p_T$ requirements for each trigger.
In the $\lblmmm$ case we consider an additional uncertainty of $6.6\%$
due to the unknown $\lbpsilm$ polarization.

For the absolute branching ratio measurements we assign the uncertainties on the world average $\br(\bpsih)$~\cite{Nakamura:2010zzi_APS} or
the most recent measurement~\cite{Abazov:2011wt_APS}.
Contributions from other sources (e.g., background PDF shape or the decay model of the simulated event) are minor ($0.3\%$--$3.4\%$).

The combined branching ratio is calculated by assuming isospin symmetry and using the $\bp$ and $\bz$ total widths~\cite{Nakamura:2010zzi_APS}.
These numbers are consistent with our previous results~\cite{Aaltonen:2008xf_APS_Aaltonen:2011cn}, 
$B$-factory measurements~\cite{Aubert:2008ps,Wei:2009zv}, and theoretical expectations~\cite{Chen:2001zc_APS,Aliev:2010uy_APS}.

\begin{table}[b]
  \caption{ Measured branching ratios of rare modes. First (second) uncertainty is statistical (systematic).
The last two values are for the isospin average.} 
  \begin{center}
    \begin{tabular}{lccc}
      \hline\hline
      Mode  & Relative $\br(10^{-3})$ & ~~~~Absolute $\br(10^{-6})$ \\ 
      \hline
      $\lblmmm$ & $\lmmmRelBr \pm \lmmmRelBrStat \pm \lmmmRelBrSyst$ & $\lmmmBr \pm \lmmmBrStat \pm \lmmmBrSyst$ \\
      $\bsphimm$ & $\phimmRelBr \pm \phimmRelBrStat \pm \phimmRelBrSyst$ & $\phimmBr \pm \phimmBrStat \pm \phimmBrSyst$ \\
      $\bpkmm$ & $\kmmRelBr \pm \kmmRelBrStat \pm \kmmRelBrSyst$       & $\kmmBr \pm \kmmBrStat \pm \kmmBrSyst$ \\
      $\bzkstmm$ & $\kstmmRelBr \pm \kstmmRelBrStat \pm \kstmmRelBrSyst$ & $\kstmmBr \pm \kstmmBrStat \pm \kstmmBrSyst$ \\
     $\bzkzmm$ & $\ksmmRelBr \pm \ksmmRelBrStat \pm \ksmmRelBrSyst$       & $\ksmmBr \pm \ksmmBrStat \pm \ksmmBrSyst$ \\
      $\bpkstpmm$ & $\kstpmmRelBr \pm \kstpmmRelBrStat \pm \kstpmmRelBrSyst$ & $\kstpmmBr \pm \kstpmmBrStat \pm \kstpmmBrSyst$ \\
     \hline
      $\bkallmm$ &                    --                                & $\kallmmBr \pm \kallmmBrStat \pm \kallmmBrSyst$ \\
      $\bkstallmm$ &                  --                                & $\kstallmmBr \pm \kstallmmBrStat \pm \kstallmmBrSyst$ \\
       \hline\hline 
    \end{tabular}
  \end{center}
  \label{tab:summary_br}
\end{table}

\begin{figure}[t]
  \begin{center}
\setlength{\tabcolsep}{1pt}
\begin{tabular}{cc}
\includegraphics[scale=0.21,clip]{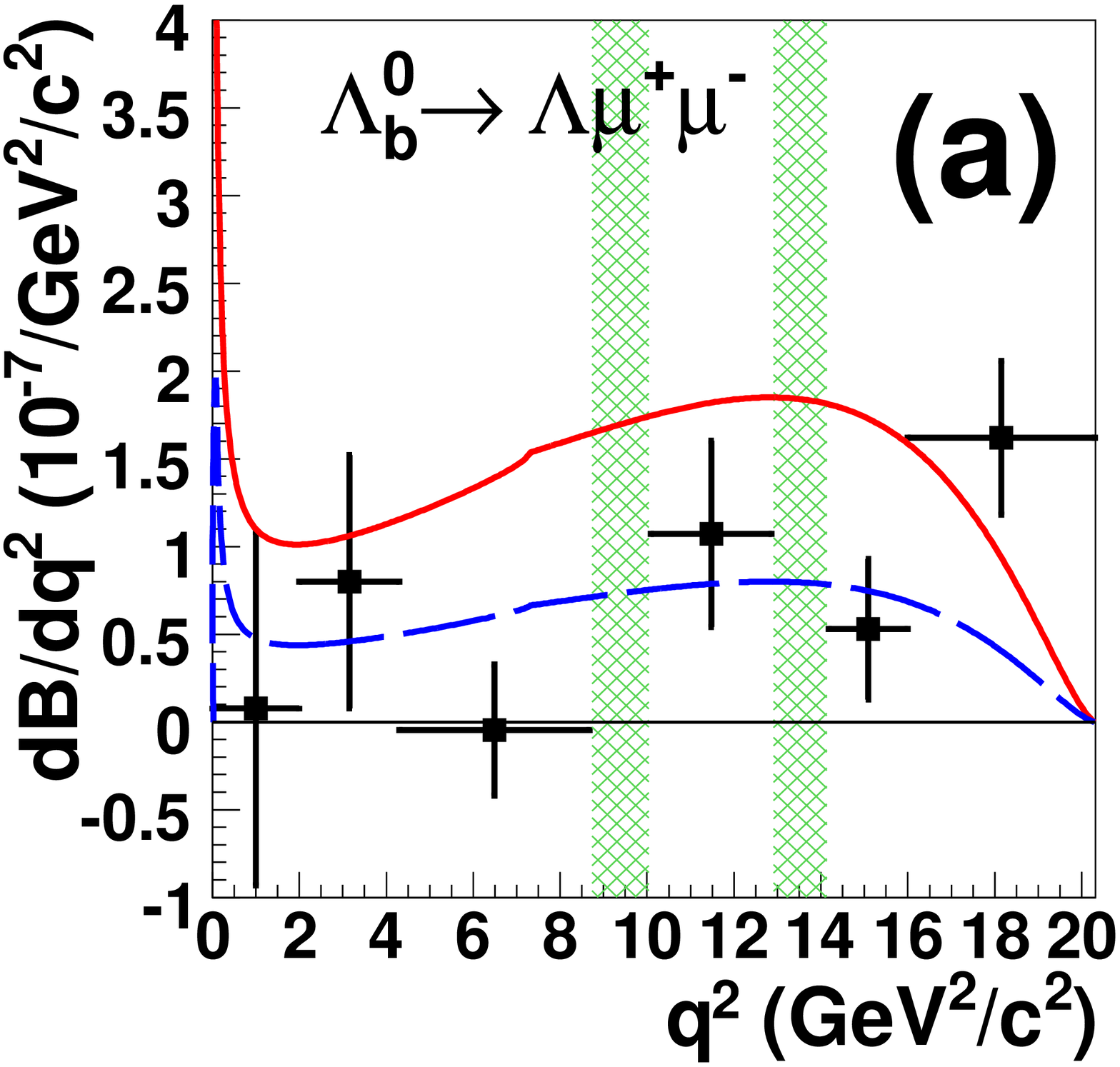}& 
\includegraphics[scale=0.21,clip]{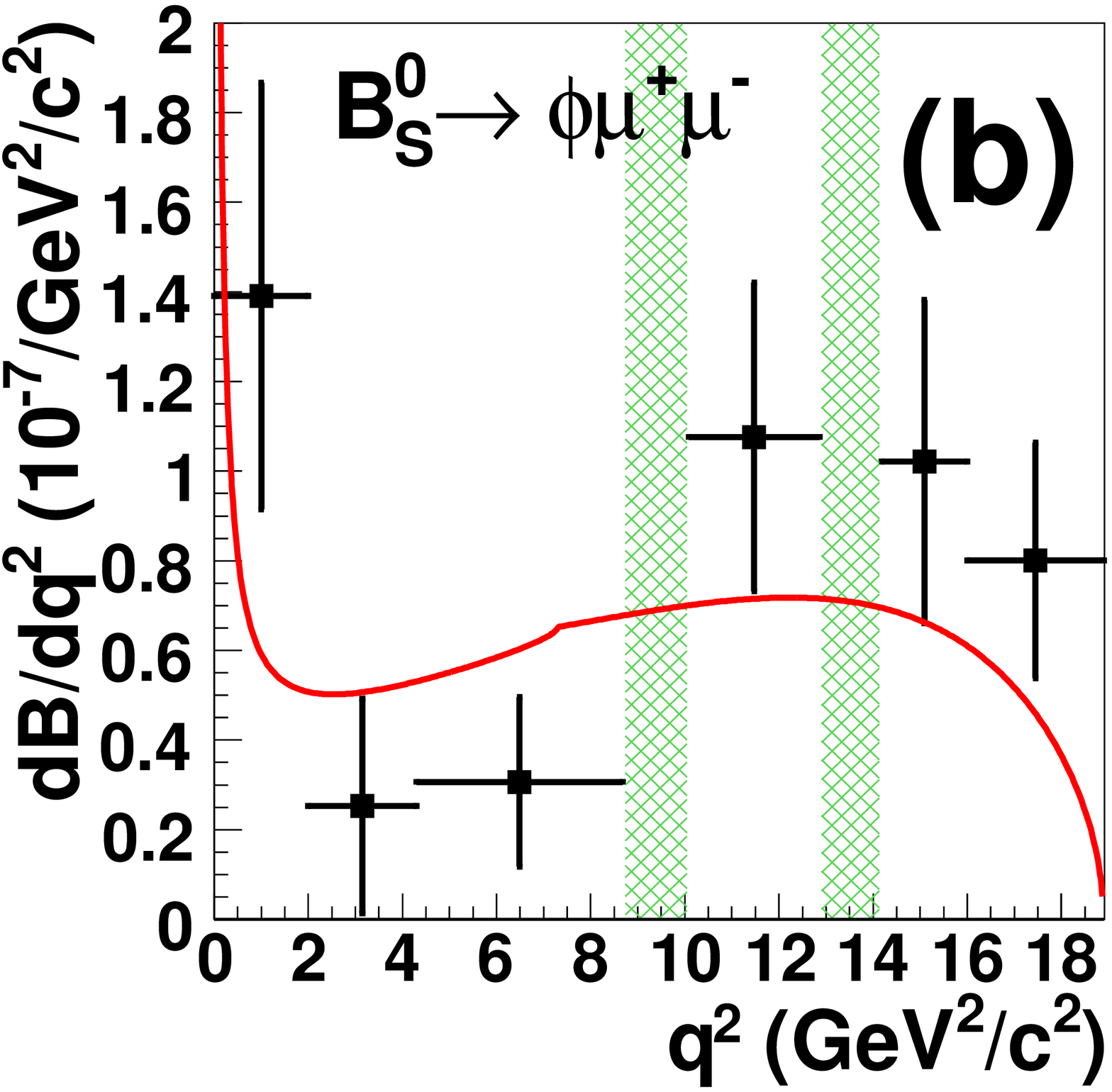} \\
\includegraphics[scale=0.21,clip]{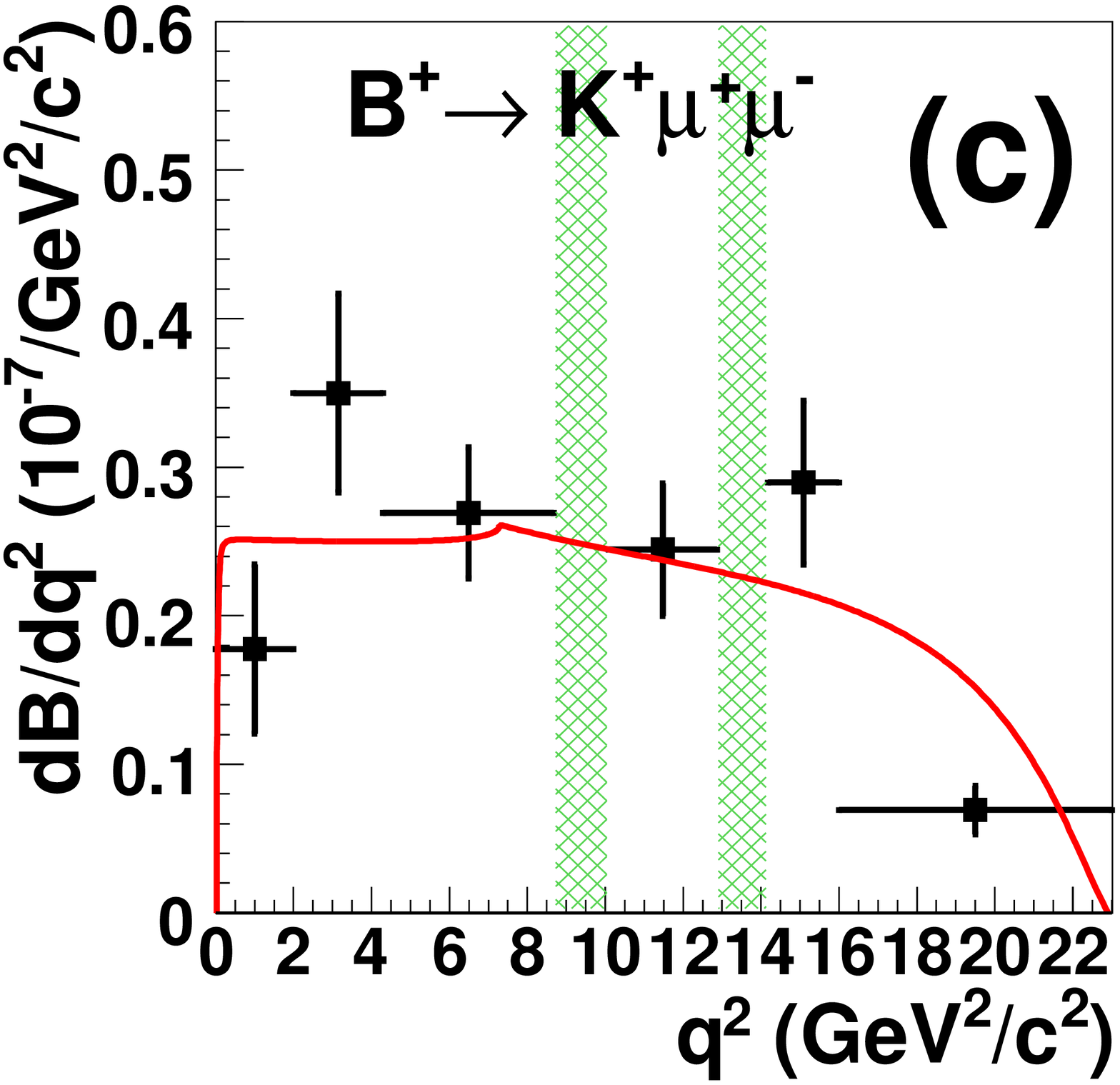}&
\includegraphics[scale=0.21,clip]{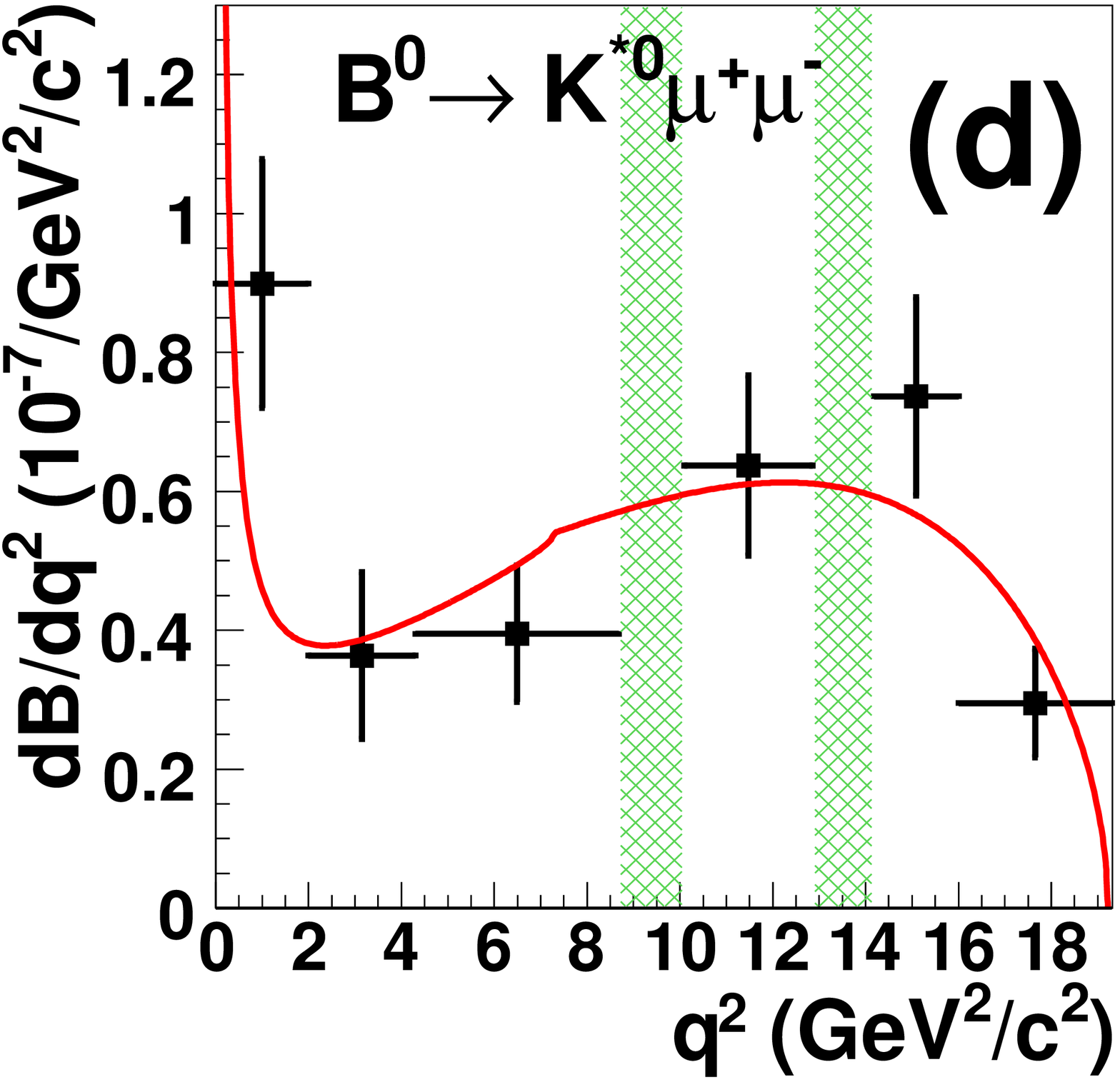} \\
    \end{tabular}
  \end{center}
  \caption{Differential branching ratios of (a) $\lblmmm$,
   (b) $\bsphimm$, (c) $\bpkmm$, and (d) $\bzkstmm$.
 The points are the fit result.
The solid curves are the SM expectation~\cite{Aliev:2010uy_APS,Ali:1999mm_APS,Ball:2004ye_Ball:2004rg_APS}.
The dashed line in the $\lblmmm$ plot is the SM prediction
 normalized to our total branching ratio measurement.
The hatched regions are the charmonium veto regions.}
  \label{fig:dbr}
\end{figure}

We also measure differential branching ratios with respect to $q^2$.
We divide the signal region
into six bins in $q^2$.
We fit the signal yield in each $q^2$ bin. In each fit, we fix the mean of the $H_b$ mass and the background slope
 to the value from the global fit, so that only the signal fraction is allowed to vary in the fit.
 Figure~\ref{fig:dbr} shows the differential branching ratios for $\lblmmm$, $\bsphimm$, and $\bkmmincl$.
For illustration, we superimpose the SM expectations, which are based on the formula in Ref.~\cite{Ali:1999mm_APS},
 with the form factors in Ref.~\cite{Ball:2004ye_Ball:2004rg_APS}, except for the case of $\lblmmm$ decays which follows Ref.~\cite{Aliev:2010uy_APS}. The cusp at $q^2\sim7\,\GeV^2/c^2$ is due to a change in parameter approximations.
Tables~\ref{tab:dbr_vs_q2_1} and ~\ref{tab:dbr_vs_q2_2} summarize the differential branching ratio measurements. The two bottom rows in each table show the results for the semi-inclusive bins which are included with ranges covering theoretically well-controlled regions.

\begin{table*}[htpb]
 \caption{Differential branching ratios of $\lblmmm$, $\bpkmm$, $\bzkzmm$, combined $\bkallmm$, 
in units of $10^{-7}$. The $q^2_{\rm max}$ is 20.30 (23.00) $\GeV^2/c^2$ for $\lmmm$ ($\kallmm$).
The first (second) uncertainty is statistical (systematic).}
  \label{tab:dbr_vs_q2_1}
  \begin{center}
    \begin{tabular}{ccccc}
      \hline
      \hline
      $q^2$ ($\GeV^2/c^2$) & $\lblmmm $ & $\bpkmm $ & $\bzkzmm $ & $\bkallmm $  \\
      \hline 
      $[0.00,2.00)$           & $0.15\pm2.01\pm0.05$  & $0.36\pm0.11\pm0.03$ & $0.312\pm0.372\pm0.024$  & $0.33\pm0.10\pm0.02$  \\
      $[2.00,4.30)$           & $1.84\pm1.66\pm0.59$  & $0.80\pm0.15\pm0.05$ & $0.929\pm0.485\pm0.070$  & $0.77\pm0.14\pm0.05$  \\
      $[4.30,8.68)$           & $-0.20\pm1.64\pm0.08$ & $1.18\pm0.19\pm0.09$ & $0.663\pm0.510\pm0.052$  & $1.05\pm0.17\pm0.07$  \\
      $[10.09,12.86)$         & $2.97\pm1.47\pm0.95$  & $0.68\pm0.12\pm0.05$ & $-0.030\pm0.223\pm0.005$ & $0.48\pm0.10\pm0.03$  \\
      $[14.18,16.00)$         & $0.96\pm0.73\pm0.31$  & $0.53\pm0.10\pm0.03$ & $0.726\pm0.257\pm0.055$  & $0.52\pm0.09\pm0.03$  \\
      $[16.00,q^2_{\rm max})$ & $6.97\pm1.88\pm2.23$  & $0.48\pm0.11\pm0.03$ & $0.214\pm0.182\pm0.016$  & $0.38\pm0.09\pm0.02$  \\
      \hline		      		     	  
      $[0.00,4.30)$           & $2.65\pm2.52\pm0.85$  & $1.13\pm0.19\pm0.08$ & $1.268\pm0.622\pm0.096$  & $1.07\pm0.17\pm0.07$  \\
      \hline		      		     	  
      $[1.00,6.00)$           & $1.27\pm2.08\pm0.41$  & $1.41\pm0.20\pm0.10$ & $0.980\pm0.614\pm0.076$  & $1.29\pm0.18\pm0.08$  \\
      \hline
      \hline
    \end{tabular}
  \end{center}
\end{table*}
\begin{table*}[htpb]
 \caption{Differential branching ratios of $\bsphimm$, $\bzkstmm$, $\bpkstpmm$, and combined $\bkstallmm$, 
in units of $10^{-7}$. The $q^2_{\rm max}$ is 18.90 (19.30) $\GeV^2/c^2$ for $\phimm$ ($\kstallmm$).
The first (second) uncertainty is statistical (systematic).}
  \label{tab:dbr_vs_q2_2}
  \begin{center}
    \begin{tabular}{ccccc}
      \hline
      \hline 
      $q^2$ ($\GeV^2/c^2$) & $\bsphimm $ & $\bzkstmm $ & $\bpkstpmm $ & $\bkstallmm $  \\
      \hline 
      $[0.00,2.00)$           & $2.78\pm0.95\pm0.89$ & $1.80\pm0.36\pm0.11$ & $1.30\pm0.98\pm0.14$ & $1.73\pm0.33\pm0.10$ \\ 
      $[2.00,4.30)$           & $0.58\pm0.55\pm0.19$ & $0.84\pm0.28\pm0.06$ & $0.71\pm1.00\pm0.15$ & $0.82\pm0.26\pm0.06$ \\ 
      $[4.30,8.68)$           & $1.34\pm0.83\pm0.43$ & $1.73\pm0.43\pm0.15$ & $1.71\pm1.58\pm0.49$ & $1.72\pm0.41\pm0.14$  \\ 
      $[10.09,12.86)$         & $2.98\pm0.95\pm0.95$ & $1.77\pm0.36\pm0.12$ & $1.97\pm0.99\pm0.22$ & $1.77\pm0.34\pm0.11$ \\ 
      $[14.18,16.00)$         & $1.86\pm0.66\pm0.59$ & $1.34\pm0.26\pm0.08$ & $0.52\pm0.61\pm0.09$ & $1.21\pm0.24\pm0.07$ \\ 
      $[16.00,q^2_{\rm max})$ & $2.32\pm0.76\pm0.74$ & $0.97\pm0.26\pm0.07$ & $1.57\pm0.96\pm0.17$ & $0.88\pm0.22\pm0.05$ \\ 
      \hline		      												  
      $[0.00,4.30)$           & $3.30\pm1.09\pm1.05$ & $2.60\pm0.45\pm0.17$ & $2.01\pm1.39\pm0.27$ & $2.53\pm0.43\pm0.15$ \\ 
      \hline		      												  
      $[1.00,6.00)$           & $1.14\pm0.79\pm0.36$ & $1.42\pm0.41\pm0.12$ & $2.57\pm1.61\pm0.40$ & $1.48\pm0.39\pm0.12$ \\  
      \hline
      \hline 
    \end{tabular}
  \end{center}
\end{table*}

In summary,
we have updated our previous analysis of the flavor-changing neutral current decays $\bsmm$ using 
data corresponding to an integrated luminosity of $\lumi$ and adding new decay channels.
We report the first observation of $\lblmmm$ and measure the
total and differential branching ratios of this decay with respect to $q^2$.
We also measure the total and differential branching ratios of $\bkmmincl$ and $\bsphimm$, with respect to $q^2$.
All measurements are consistent and competitive with other results, 
and the differential measurements of $\bsphimm$ and $\lblmmm$ are the first such measurements.
At present there is no evidence of discrepancy from the SM prediction.

\begin{acknowledgments}
We thank the Fermilab staff and the technical staffs of the participating institutions for their vital contributions. This work was supported by the U.S. Department of Energy and National Science Foundation; the Italian Istituto Nazionale di Fisica Nucleare; the Ministry of Education, Culture, Sports, Science and Technology of Japan; the Natural Sciences and Engineering Research Council of Canada; the National Science Council of the Republic of China; the Swiss National Science Foundation; the A.P. Sloan Foundation; the Bundesministerium f\"ur Bildung und Forschung, Germany; the World Class University Program, the National Research Foundation of Korea; the Science and Technology Facilities Council and the Royal Society, UK; the Institut National de Physique Nucleaire et Physique des Particules/CNRS; the Russian Foundation for Basic Research; the Ministerio de Ciencia e Innovaci\'{o}n, and Programa Consolider-Ingenio 2010, Spain; the Slovak R\&D Agency; and the Academy of Finland. 
\end{acknowledgments}
\bibliographystyle{h-physrev5_cdf_prl}
\bibliography{reference}

\begin{thebibliography}{10}
\bibitem{CDF_coordinate}
We use a cylindrical coordinate system in which $\theta$ is the polar angle
  with respect to the proton beamline and pseudorapidity $\eta \equiv -\ln(\tan
  \theta/2)$.
\end{thebibliography}

\end{document}